# To Learn or Not to Learn:
# Deep Learning Assisted Wireless Modem Design

XUE Songyan[1], LI Ang[1], WANG Jinfei[1], YI Na[1], MA Yi[1*], TAFAZOLLI Rahim[1], and DODGSON Terence[2]
([1]Institute for Communication Systems, University of Surrey, Guildford, GU2 7XH, United Kingdom)
([2]Airbus Defense and Space, Portsmouth, PO3 5PU, United Kingdom)
[*]Correspondence author: y.ma@surrey.ac.uk

**Abstract**
**Deep learning is driving a radical paradigm shift in wireless communications, all the way from the application layer down to the physical layer. Despite this, there is an ongoing debate as to what additional values artificial intelligence (or machine learning) could bring to us, particularly on the physical layer design; and what penalties there may have? These questions motivate a fundamental rethinking of the wireless modem design in the artificial intelligence era. Through several physical-layer case studies, we argue for a significant role that machine learning could play, for instance in parallel error-control coding and decoding, channel equalization, interference cancellation, as well as multiuser and multiantenna detection. In addition, we will also discuss the fundamental bottlenecks of machine learning as well as their potential solutions in this paper.**

**Keywords**
**Deep learning, neural networks, machine learning, modulation and coding.**

## 1 Introduction

With the launch of commercial 5G mobile networks in 2019, wireless communications research is now well on the way towards Vision 2030 and beyond. Today, the picture of future wireless communications is becoming much clearer than ever. According to ITU Network 2030 Working Group [1], future networks should be architected to support holographic communications and smart connectivity, providing seemingly zero latency, guaranteed ultra-reliability (e.g. 99.9999%), massive IoT connectivity, and Tbit/s wireless speed. Communication networks are no longer only a medium for information flow, but also act as distributed computer to form over-the-top (OTT)-like platform to provide services (such as computing-as-a-service, design-as-a-service, etc.) for vertical users. To achieve this goal, wireless technologies should be fundamentally re-designed to be able to explore fully of the spectrum; as such this is driving the development of extreme physical-layer (PHY) technologies, which are able to handle wireless systems with many nonlinearities, due to the use of very-high order modulations, unexploited mmWave or THz bands, and/or low-cost electronic components (such as low-noise amplifiers, LNAs, mixers, oscillators and low-resolution analog-to-digital converters, ADCs). Moreover, PHY solutions should be made scalable to the number of connected devices; and they should be parallel computing ready, as future high-performance computing technologies (including future quantum computing technology) rely highly on the parallel computing power.

With such a big picture in mind, machine learning or more specifically, deep learning can play a significant role in the PHY design, at least from the following five aspects:

1) Conventional PHY algorithms, particularly for wireless receivers, are mostly not parallel computing ready. For instance, most of the linear or nonlinear coherent receivers (such as linear zero-forcing, minimum mean-square error, lattice reduction, sphere decoding) require either channel matrix inversions or channel matrix decompositions, which are difficult to execute in an efficient and parallel manner. This can cause a bottleneck for the implementation of advanced channel equalizers or multiuser detectors at the receiver side. An exception could be the matched-filter algorithm, which is of low-complexity and parallel computing architecture. On the other hand, matched filtering is often too suboptimum for most wireless applications. One might also argue for parallel computing abilities of brute-force search, likelihood ascent search, or Tabu search. However, those algorithms trade off complexity for parallel computing, and thus they are not cost-effective solutions. In this paper, we will study the merits of deep-learning assisted solutions, with specific to their inborn parallel computing ability.
2) Conventional hand-engineered PHY algorithms face the fundamental trade-off between performances and complexities. Optimum algorithms are often too complex to implement, and low-complexity algorithms are often too suboptimum.

[1] This work was supported in part by EU H2020 5G-DRIVE Programme (Grant: 814956), in part by Airbus Defense and Space, and in part by the UK 5G Innovation Centre (5GIC).

Deep-learning assisted PHY algorithms have the potential to achieve (near-)optimum performances with low computation complexities. We argue for the merits of performance-complexity trade-off when using deep learning.

3) Current PHY technologies are designed for linear communication channels, and they are not optimized for future wireless systems often operating in nonlinear conditions. Nonlinear systems are often much harder for mathematical analysis, and in general, we even do not know their channel capacities. Hand-engineered approaches for PHY design and optimization are currently very challenging; and this is where deep learning can be of much assistance.

4) Sensing and communication is an emerging concept in the scope of network automation. Basically, wireless networks are able to capture environmental changes through local and remote sensors or even live video records, based on which networks can adapt their operating states for optimum uses of their local radio resources. On the PHY layer, environmental information can be translated into channel side information through machine learning [2], and this can be useful for advanced modem functions such as adaptive modulation, coding and beamforming. In addition, machine learning can play a central role in building and reconfiguring state machines for local networks through extensive online background learning.

5) Since Shannon's ground-breaking work on communication theory reported in 1948, most telecommunications research effort has been targeting the Level A problem, i.e., how accurately can the information-bearing symbols be conveyed from one point to another? In the academic domain, this research problem has been almost saturated. In the industrial domain, it is very challenging to apply the outcome of Level A research so as to satisfy the growing demand of future wireless networks in terms of smart connectivity, providing seemingly zero latency and perceived infinite capacity. Therefore, it is perhaps the right time to revisit or invest more research effort on the Level B problem, i.e., how precisely do the symbols of communication convey the desired meaning? This problem goes well beyond traditional source encoding practices, as for now source encoders are expected to understand the meaning of objects instead of just the probability distribution. A simple example of the Level B problem is illustrated in Figure 1, where the picture on the left-hand side is the original picture for transmission. Instead of compressing the picture using current codec processing methods, source encoders that have been trained to understand the meaning of the picture could send a textual description such as, "a white background picture, with a mother kangaroo carrying her baby in her pouch." The receiver then rebuilds the picture based on the meaning of the received symbols; This can be termed semantic communications, which involves heavy use of artificial intelligence/machine learning in semantic source encoding and decoding.

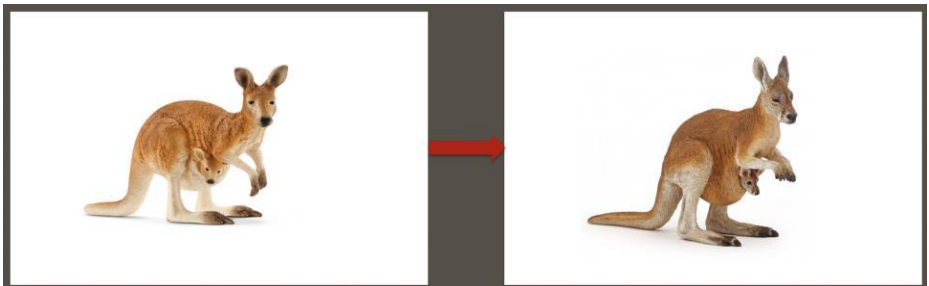

Figure 1. A simple example of Level B communication problem (Semantic Communication)

Certainly, we shall be able to find more merits and interesting topics when applying artificial intelligence/machine learning in wireless communications; some are already under fast development, and some are just emerging. In the following sections, our discussion will be mainly focused on points 1), 2), and 3), as they are suitable for both current and future communication networks. In addition, we will also discuss fundamental bottlenecks when applying deep learning to wireless modem design.

The rest of this paper is organized as follows. Section II outlines the principles of deep learning assisted modem design in wireless communication physical layer. Section III provides the design details of three practical physical layer applications. Section IV provides further discussions and open research problems. Section VI draws the conclusion.

## 2 Principles of Deep Learning Assisted Modem Design

By deep learning, we often mean machine learning through deep artificial neural networks (ANNs). An ANN is called deep when it has two or more hidden layers. Mathematically, the main function of each hidden layer is to perform classification of input vectors which might be referred to as perception in the artificial intelligence domain. If each output neuron yields a binary-type output, a hidden layer, consisting of $L$ neurons, is able to classify at least $L$ clusters. When a hidden layer is trained according to the nearest-neighbor rule, the machine is able to learn optimum classifications [3]. One might also employ the k-nearest neighbor rule to train the hidden layer, and in this case, the machine can form at most $2^L$ clusters. This is a possible way to scale up ANN when input vectors have to be partitioned into clusters that are growing exponentially. However, we will have to trade off the classification accuracy.

Prior to studying deep learning assisted wireless modem design, let us have a brief review of the PHY procedure of point-to-point communications; as illustrated in Figure 2. Basically, signal waveforms are drawn from a finite-alphabet set, say $A$, with the size $J$. After going through the fading channel, received waveforms in their discrete-time equivalent form are vectors forming an infinite set. The role of receivers is to map the received vectors back onto the finite-alphabet set $A$. This procedure mimics the ANN-based classification procedure; as described above. Indeed, it is rather straightforward to replace the receiver box in Figure 2 with an ANN black-box. The input vectors are formed by received waveforms combined with the channel state information, as they together form a bijection to the original waveform set $A$. Alternatively, the input vectors can be channel-equalized signals which also form bijection with the original waveform set $A$. The bijection allows the ANN black-box to be trained through supervised learning. In fact, this example is not the only way to apply deep learning for modem designs. It is also possible to replace both the transmitter and receiver with their corresponding ANN black-boxes, so as to form an autoencoder which can be trained end-to-end for joint transmitter and receiver design [4, 5]. Theoretically, a shallow-ANN (i.e. an ANN with a single hidden layer) would be sufficient to perform signal classification at the receiver side, as a receiver is normally a single-task classifier. Joint transmitter and receiver designs (autoencoders) are different, as they need at least one hidden layer at the transmitter side to construct the waveform set and another hidden layer at the receiver side to carry out corresponding signal classification. Here, the implication is that deep-ANN is more meaningful when a PHY module or procedure can have a breakdown of two or more different tasks; or otherwise, a shallow-ANN would be more than enough. This issue will be further elaborated in Section 3.

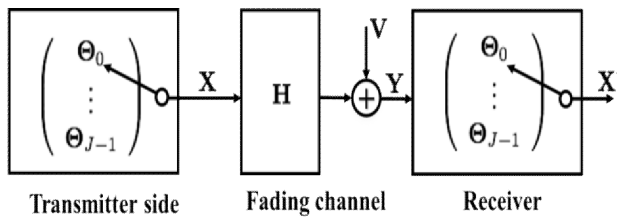

Figure 2. Block diagram of the PHY procedure of point-to-point communication.

In addition to the ANN architecture, ANN training algorithms or methods are crucial when improving machine learning efficiency. Analogous to ANN-assisted machine learning practices in the general artificial intelligence domain, it is always important to pay particular attention to the following three aspects:

1) Weighting vectors (including biases) in each hidden layer should be carefully initialized. They are often randomly generated according to a certain independent probability distribution within a certain range, which can vary from case to case in practical applications. Specific to modem design, we should bear in mind that those weighting vectors during training become reference vectors for the eventual signal classification. Therefore, they should be initialized in a way that facilitates the capture of the characteristics of communication signals by machines.
2) Activation functions must be carefully selected to improve the optimality or efficiency of ANN-assisted machine learning. For instance, Softmax(.) is suitable for small-scale ANNs to adopt the nearest-neighbor rule in machine learning. This enables Euclidean-distance optimality when training a hidden layer. Moreover, Softmax(.) allows machines to produce soft outputs that are often useful for soft-demodulation and decoding practices. Alternatively, we can employ Sigmoid(.) to scale up ANNs when they are expected to handle massive-region classifications. Certainly, we will have to pay for the classification optimality. For more information, a relatively comprehensive list of activation functions as well as their descriptions can be found in [6].
3) Backpropagation (BP) is essential at the ANN training stage to recursively update neuron weighting vectors, with the aim of minimizing the loss function such as the mean-square error, mean absolute error or categorical cross-entropy between the ANN output and labeled training target, depending on the applications. A commonly used BP method is called mini-batch gradient descent, which randomly picks up a certain number of training samples from the entire training data set on each training iteration. Compared to another commonly used BP algorithm called batch gradient descent, mini-batch gradient descent can significantly reduce computational complexities, particularly when the path to the desired minima is quite noisy.

## 3 Deep Learning Assisted Modem Designs and Their Merits

In this section, we will offer three case studies on deep-learning assisted wireless modem design, and argue for their advantages in computing latency reduction, remarkable complexity-performance trade-off as well as robustness to nonlinear physical distortions.

**Case Study 1: Deep Learning Assisted Parallel Decoding of Convolutional Codes**

Error-control codes often have a serial computing architecture in nature due to correlations amongst codeword bits. This fact is challenging the design of parallel-computing ready decoding algorithms. Recent advances towards ANN-assisted decoders are mainly based on recurrent neural networks [7, 8]; and there is a clear show of advantages in performance-complexity trade-off. Here, we review a more recent contribution in this domain, which proposes to the employment of feed-forward neural networks for low-complexity parallel decoding of convolutional codes [9].

The basic idea is to partition a long convolutional codeword into a number of pieces, forming so-called sub-codewords. When the length of sub-codewords is sufficiently long, there exists a bijection between sub-codewords and their corresponding original information bits, subject to an initial state uncertainty. As depicted in Figure 3-(a), sub-codewords are first decoded in parallel using a list maximum-likelihood decoder (List-MLD), and then initial state uncertainties are removed through the sub-codeword merging process; referred to as a two-stage decoding process that can be implemented in parallel. In this case study, the role of the ANN is to replace the List-MLD algorithm at the sub-codeword decoding stage, as the latter is of very high computation complexity. Figure 3-(b) illustrates the ANN training procedure, where the sub-codeword decoder is modelled as a deep-ANN black-box. The input vector is the noisy version of all possible sub-codewords, and the output vector is the corresponding estimate of the original information bits. It is worthwhile highlighting that the training set of input vectors should be carefully defined so as to incorporate the effect of initial state uncertainty (as detailed in [9]), as this is crucial for the sub-codeword merging stage. Moreover, it is suggested to partition a long convolutional codeword evenly, as in this case we only need to train one ANN block-box and can reuse it for all sub-codewords; thus, resulting in an efficient way to reduce the training complexity.

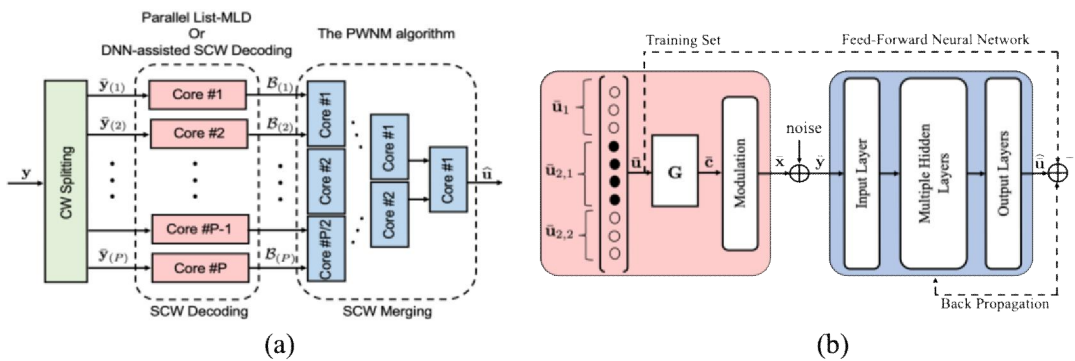

Figure 3. Illustration of the ANN-assisted parallel decoder: (a) two-stage parallel decoding, and (b) ANN-assisted sub-codeword decoder. There are three hidden layers, with each employing ReLU activation function, the output layer is equipped with sigmoid activation function which outputs the estimated original information bits.

Figure 4 illustrates the bit reliability of convolutional decoders in additive white Gaussian noise (AWGN), considering a ½-rate non-recursive convolutional code with a codeword length of 64. The illustrated simulation results are only for Eb/No=4 dB; and similar conclusions can be drawn for other Eb/No's; please see [9] for details. The ANN black-box was trained at Eb/No=2 dB. When comparing the parallel decoder with the conventional MLD, it can be seen from Figure 4 that they have no difference in bit reliability; and thus, the parallel decoder is optimum. Moreover, due to the parallel computing nature, the parallel decoder has the potential to reduce computing latency, subject to the number of sub-codewords. When the sub-codeword decoder is realized through the ANN black-box described in Figure 3-(b), we can see a little bit of a performance loss in bit reliability (around 0.03%); this is mainly due to using an insufficient number of epochs during the ANN training stage. Nevertheless, the computation time for sub-codeword decoding is reduced by around 95%. It is clear that ANN helps to achieve a very good complexity-performance trade-off. In addition, the ANN decoder can be executed fully in parallel; and this is an additional advantage for latency reduction.

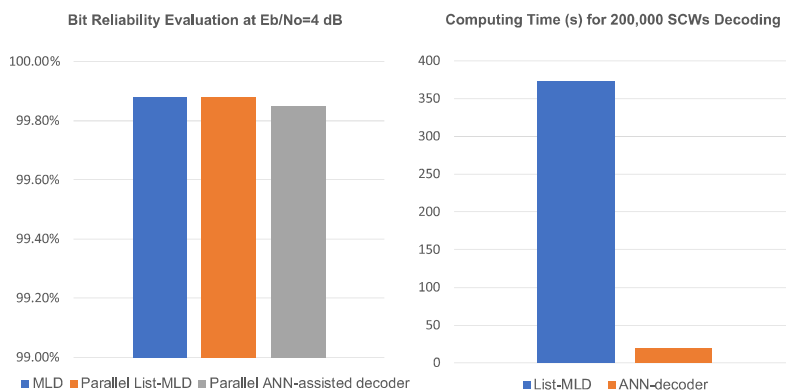

Figure 4. Bit reliability and latency evaluation for decoding of ½-rate non-recursive convolutional codes with a codeword length of 64.

**Case Study 2: Deep Learning Assisted Multiuser OFDMA Frequency Synchronization**

Consider a multiuser frequency-synchronization problem in the context of orthogonal frequency-division multiple-access (OFDMA) uplink communications, where transmitters experience independently generated carrier-frequency-offsets (CFOs), due to oscillator instability or Doppler-induced random frequency modulations. This problem involves two sub-problems. One is the multiuser-CFO estimation, and the other is multiuser detection (MUD) or multiuser interference (MUI) cancellation given the CFO estimates. Multiuser-CFO estimation can be implemented by employing either pilot-assisted approaches or blind approaches that exploit statistical behaviors inherent in signal waveforms. When CFO estimates are assumed available at the transmitter side, each transmitter can carry out CFO pre-compensation, individually. However, link-level latency will be a considerable issue due to the CFO feedback delay. Alternatively, multiuser frequency synchronization can also be carried out at each individual user domain (e.g., sub-band) using the filterbank approach, which can be combined with iterative parallel interference cancellation (PIC). However, such a method is vulnerable to the CFO estimation accuracy, and it could introduce extra baseband processing latency into the system.

Figure 5 illustrates a deep-learning assisted multiuser frequency synchronization approach, named classification-and-then-MUD (CAT-MUD) in [10]. The deep-ANN has two functional layers: one is responsible for multiuser-CFO classification, and the other is for the MUI cancellation. The CFO classifier is employed to tell the CFO sub-range where transmitters' CFOs fall in. This is very different from the conventional CFO estimation in the sense that the classifier only estimates the CFO range instead of CFOs. With the estimated CFO sub-range index, received signals are then fed into the MUD layer for the MUI cancellation; please find a detailed introduction of CAT-MUD in [10].

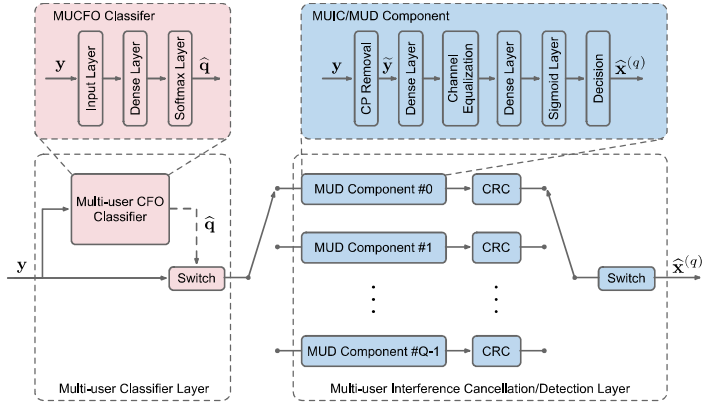

Figure 5. Block diagram of the deep-ANN assisted CAT-MUD.

Figure 6 illustrates the overall system performance (in block-error rate, BLER) for OFDMA systems, where 4 transmitters evenly share 32 subcarriers. Original information bits are first modulated into 16-QAM symbols and then transmitted through an 8-tap frequency-selective Rayleigh fading channel (3GPP Channel Model A). To be more robust to CFO classification errors, the switch depicted in Figure 5 can simultaneously turn on multiple adjacent MUD branches. Figure 6 shows that the 3-branch model achieves the best performance-complexity trade-off. It outperforms the conventional PIC approach by around 3 dB in Eb/No and offers comparable performances with the CFO-free case at low and moderate SNRs (such as Eb/No<15 dB).

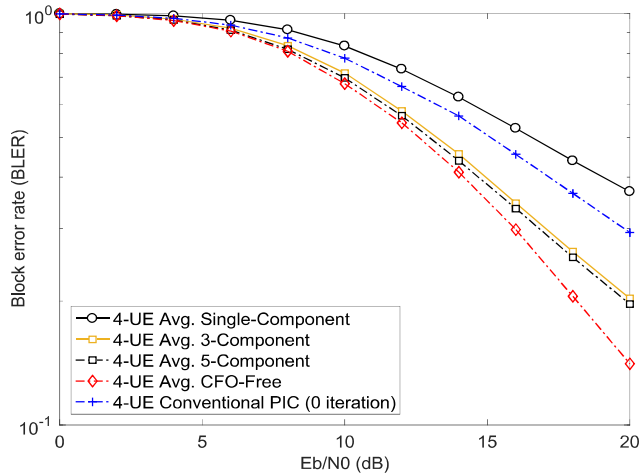

Figure 6. BLER of CAT-MUD as a function of Eb/No (dB) over Rayleigh fading channels.

**Case Study 3: Deep Learning Assisted Coherent MIMO Detection**

Multiuser multiple-input multiple-out (MU-MIMO) signal detection over noisy fading channel is mathematically an integer least-squares (ILS) problem, which aims to minimize the pairwise Euclidean distance between the transmitted signal multiplied by channel matrix and the received signal [11]. Concerning the optimal MLD solution to be computationally too expensive, the usual practice is to employ linear channel equalization algorithms such as the matched filter (MF), zero forcing (ZF) and linear minimum mean-square error (LMMSE) to trade off the optimality for lower computational complexity. However, linear algorithms are often too sub-optimum due to their use of symbol-by-symbol detection. Therefore, enormous research efforts have been paid in the last two decades to achieve the best performance-complexity trade-off through the use of non-linear algorithms (e.g., V-BLAST [12], LMMSE-SIC [13] and so on). The problem is that most of the non-linear algorithms are too complex for current DSP technology and do not lend themselves well to parallel computing. This goes against the trend of computing technology development.

Deep learning assisted solutions have demonstrated their potential for offering computational complexity close to linear receivers, without compromising the detection performance. Moreover, most of the deep learning algorithms are parallel computing ready. According to the ways of utilizing channel state information at the receiver side (CSIR), deep-learning solutions can be divided into two categories: channel equalization and learning (CE-L) mode (as shown in Figure 7-a) and direct learning (Direct-L) mode (as shown in Figure 7-b). The difference is that the CE-L mode employs ANN black-box after channel equalization, and the Direct-L mode takes both CSIR and received signal as the input vector for signal classification.

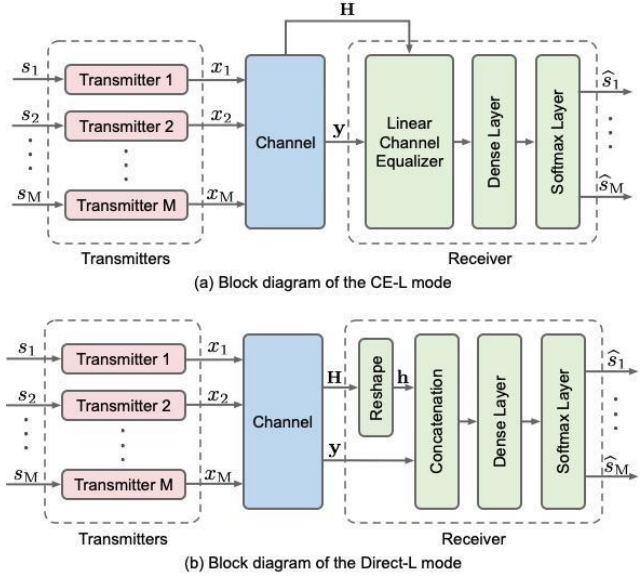

Figure 7. Block diagram of deep-learning assisted MU-MIMO detection algorithms

A major advantage of the CE-L mode lies in the use of channel equalization for multiuser signal orthogonalization. Hence, the input vector to the ANN black-box is effectively a noisy version of the transmitted signal vector. By such means, the CE-L mode can turn the ANN classification problem from the vector level to the symbol level. However, the performance of the CE-L mode is limited by the symbol-by-symbol MLD bound. Theoretically, the Direct-L mode is able to achieve the optimum MLD performance for the vector-level classification. In addition, the Direct-L mode does not need channel equalization. This is a remarkable advantage as channel equalizers often require channel matrix inversions which do not support parallel computing. On the other hand, the Direct-L model is not a scalable approach with the size of MIMO, due to ANN's reduced classification ability with the growth of multiuser interferences.

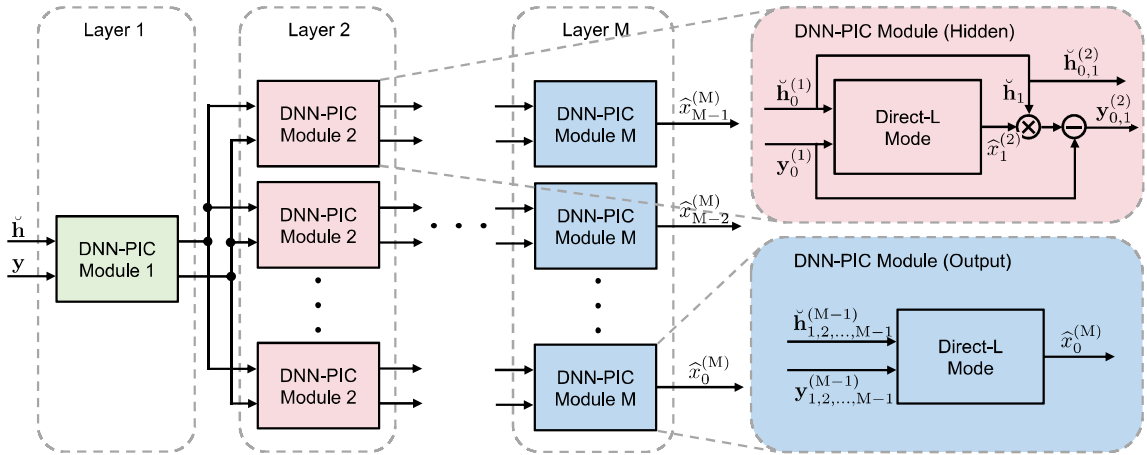

Figure 8. Block diagram of the DNN-PIC approach

Figure 8 illustrates a novel deep-ANN approach, where a multi-layer modularized ANN is combined with PIC to scale up the Direct-L mode. This approach is called DNN-PIC in [14]. Basically, the entire ANN consists of a number of cascaded PIC layers, with each layer employing a group of identical pre-trained DNN-PIC modules for signal classification and interference cancellation. Therefore, multiuser interference decreases linearly with the feed-forward procedure, and the last layer is able to provide a better classification of MU-MIMO signals.

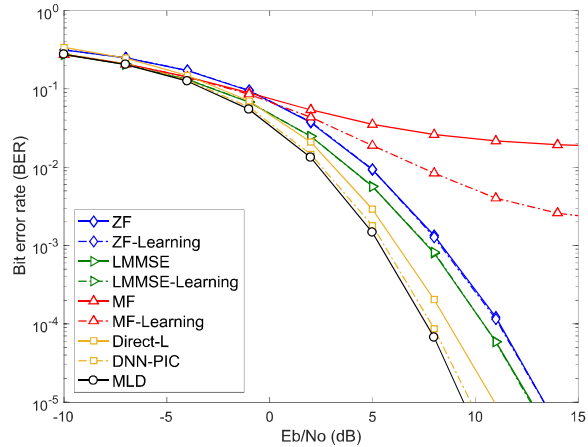

Figure 9. Average BER as a function of Eb/No for uncoded 4-by-8 MU-SIMO system with BPSK modulation.

Figure 9 compares the average bit-error-rate (BER) performance between conventional MU-MIMO receivers and the DNN-assisted solutions. For the CE-L and Direct-L mode, the ANN was trained at Eb/No=5 dB. For the DNN-PIC approach, the ANN was trained at three different Eb/No points (i.e. Eb/No=0 dB, 5 dB and 10 dB), and were optimally selected in the communication procedure in order to obtain the best achievable performance. Simulation results show that deep learning modules largely improve the detection performance of the MF-based receiver (around 8 dB at BER of $10^{-3}$) due to better use of the sequence-detection gain. For both the ZF and LMMSE receiver, the sequence-detection gain vanishes since channel equalization orthogonalizes multiuser signals. Meanwhile, the Direct-L mode significantly outperforms all CE-L modes, and this result confirms the accuracy of the theoretical analysis. Finally, the proposed DNN-PIC approach further improves the BER performance of the Direct-L approach by around 1.5 dB. The performance gap between the DNN-PIC and the MLD receiver is only about 0.2 dB. Again, it should be emphasized that the DNN-PIC approach is parallel computing ready.

## 4 Discussion and Research Challenges

Although deep learning has achieved widespread empirical success in many areas, the applications of deep learning for wireless communication physical layer design are still at the early stage of research and engineering implementation. In this section, we list several fundamental bottlenecks together with the potential future research directions.

*A. Training set overfitting*

Overfitting is a modeling problem which occurs when a function is too closely fit a limited data set [15]. In PHY, it could refer to the case that an ANN-assisted receiver trained for a specific wireless environment (or channel model) is not suitable for another environment (or channel model). It is a severe problem since a deep learning solution with limited generalization capability is less useful in real practice. However, this issue can be viewed more positively if deep learning algorithm can be used to optimize wireless receivers integrated into access points based on their local environments.

*B. Scalability of DL-based solutions*

In machine learning theory, scalability refers to the effect of an increasing training data set on the computational complexity of a learning algorithm. For instance, the ANN solution in Fig. 7(b) has its learning capacity rapidly degraded with the growth of transmit antennas [14]. The current approach to mitigate this problem is by means of training the ANN with channel equalized signals (as shown in Fig. 7(a)). However, in this case, ANN-assisted receivers are not able to exploit maximally the spatial diversity-gain due to the multiuser orthogonalization enabled by channel equalizers, and the performance goes far from optimum. To tackle this issue, novel deep learning algorithms with good scalability are required (and expected) in the future.

*C. Training strategies and performance evaluation*

Deep learning for wireless communication is a new research area, people lack experience in training strategies. For example, the optimal training SNR points for different PHY scenarios remain unknown [15]. In [9], it can be observed that the training of an ANN at relatively high SNRs gives an excellent generalization performance at low SNR regime in AWGN channel. However, when wireless channel becomes fading [14], the learned PHY feature at high SNR regime can no longer indicate the feature of low SNRs. A potential solution is to train ANNs at different SNR regimes separately and then merge the results together, but this solution introduces additional training complexities and requires SNR estimation. A related question is whether there is a more appropriate way to measure the training process in PHY solutions. It is well known that ANN training aims to minimize a given loss function, and we consider that an ANN is well trained if the loss is converged to an ideal state. On the other hand, PHY performance is normally measured by BER or SER. In most of the ANN-assisted PHY solutions, we make a hard decision on ANN outputs to obtain the bit-level (or symbol-level) estimates. However, the loss function might not be able to accurately indicate the training progress when complicated PHY scenarios are considered (e.g., high-order modulation and fast fading channel). In [14], the authors introduce a method which measures the training progress by computing the average BER/SER over the last few training epochs, and the estimated BER performance is shown to be very close to the validation performance. In general, the training strategies, especially for PHY applications, are worthy of investigation in future research.

*D. Hardware implementation*

Currently, most of the ANN-assisted PHY solutions are still in their software simulation stages, but hardware implementation normally requires more practical considerations [17, 18]. Apart from the channel model and data set that we have discussed in previous sections, power consumption also needs to be considered since the ANN training process often involves very expensive computation cost. The aim of reducing ANN learning expenses has recently motivated a new research area on *non von Neumann* computing architectures.

# 5 Conclusions

This paper presents several promising ANN-assisted PHY applications. The idea lies in the use of ANNs to replace parts of the conventional signal processing blocks in the communication chain. It is shown that ANN-assisted approaches achieve competitive performance in terms of both reliability and latency in various applications. More importantly, deep learning offers us a fundamentally new way to design and optimize the conventional communication systems. A wide range of open challenges need to be solved, and theoretical analysis is also expected in future research.